\documentclass[11pt,a4paper]{article}
\usepackage[T1]{fontenc}
\usepackage[utf8]{inputenc}
\usepackage{lmodern}
\usepackage[a4paper,margin=2.5cm]{geometry}
\usepackage{microtype}
\usepackage{authblk}
\usepackage[numbers,sort&compress]{natbib}
\usepackage{bm,latexsym,amsmath,amssymb,amsfonts,mathtools,mathrsfs}
\usepackage{nicefrac}
\usepackage{graphicx}
\usepackage{wrapfig}

\usepackage[hang]{subfigure}
\usepackage[dvipsnames]{xcolor}

\usepackage[normalem]{ulem} 
\usepackage[export]{adjustbox}

\usepackage{todonotes}
\definecolor{palatinate}{RGB}{128, 49, 123}
\definecolor{nicered}{rgb}{0.7,0.1,0.1}
\definecolor{nicegreen}{rgb}{0.1,0.5,0.1}

\newcommand{\be}{\begin{equation}}
\newcommand{\ee}{\end{equation}}
\newcommand{\beal}{\begin{aligned}}
\newcommand{\eeal}{\end{aligned}}
\newcommand\bea{\begin{eqnarray}}
\newcommand\eea{\end{eqnarray}}
\newcommand{\bec}{\begin{cases}}
\newcommand{\eec}{\end{cases}}

\numberwithin{equation}{section}

\usepackage[
  colorlinks=true,
  linkcolor=blue,
  citecolor=blue,
  urlcolor=blue,
  pdfauthor={Gabriel Arenas-Henriquez, Rodrigo Olea, David Rivera-Betancour},
  pdftitle={Regular/Singular Black Hole Thermodynamics from Renormalised Volume/Area}
]{hyperref}

\title{Regular/Singular Black Hole Thermodynamics\\
from Renormalised Volume/Area}
\author[$\alpha$]{Gabriel Arenas-Henriquez\thanks{\href{mailto:arenas@ucas.ac.cn}{\nolinkurl{arenas@ucas.ac.cn}}}}
\author[$\beta$]{Rodrigo Olea\thanks{\href{mailto:rodrigo_olea_a@yahoo.co.uk}{\nolinkurl{rodrigo_olea_a@yahoo.co.uk}}}}
\author[$\gamma$]{David Rivera-Betancour\thanks{\href{mailto:driverabetancour@itmp.msu.ru}{\nolinkurl{driverabetancour@itmp.msu.ru}}}}
\affil[$\alpha$]{School of Fundamental Physics and Mathematical Sciences, Hangzhou Institute for Advanced Study, UCAS, Hangzhou 310024, China}
\affil[$\beta$]{Instituto de F\'isica, Pontificia Universidad Cat\'olica de Valpara\'iso, Casilla 4059, Valpara\'iso, Chile}
\affil[$\gamma$]{Institute for Theoretical and Mathematical Physics, Lomonosov Moscow State University, 119991 Moscow, Russia}
\date{}

\begin{document}

\maketitle
\begin{abstract}
We study the finite mismatch between the holographically and topologically renormalised Euclidean actions of the four-dimensional AdS C-metric. Starting from the relation between Renormalised Volume and the Einstein sector of Conformal Gravity, we separate the regular curvature contribution from the contributions supported on the two conical axes. After introducing the usual Euler normalisation for smooth manifolds, the residual difference between the regular action evaluations is determined by the tension-weighted Euler terms of the renormalised worldsheet area. We identify this difference with the Gauss--Bonnet entropy contribution at the polar conical points of the horizon. The ordinary Euler characteristic remains fixed, while the deficits determine how its curvature representation is divided between regular and singular contributions. At fixed total string tension, the difference of the renormalised worldsheet areas per unit Euclidean time determines the relative tension work, giving a geometric interpretation of the relative thermodynamic length.
\end{abstract}

\tableofcontents
\section{Introduction}

The notion of \emph{Renormalised Volume} arises naturally in differential geometry when studying noncompact manifolds whose ordinary Riemannian volume is infinite, but whose asymptotic geometry is sufficiently controlled to extract a finite, geometrically meaningful quantity.

The principal setting is that of asymptotically hyperbolic manifolds, particularly Poincar\'e--Einstein manifolds, whose metrics approach hyperbolic space near a conformal boundary. Although the volume diverges, its asymptotic expansion contains a finite part that can be interpreted as a renormalised geometric invariant (with qualifications depending on the dimension and the choice of boundary representative).

The construction connects conformal geometry, Einstein manifolds, Chern--Gauss--Bonnet formulas, holographic renormalisation, conformal anomalies, and renormalised areas of minimal surfaces.

In the mathematical literature, the concept of renormalised Volume $V_{\rm{ren}}$ for asymptotically hyperbolic spaces is closely related to the existence of bulk conformal invariants in even-dimensional manifolds \cite{Graham:1999jg}. Indeed, in four spacetime dimensions, a theorem due to Anderson \cite{Anderson2000L2CA} states that the renormalised Volume is proportional to the integral of a Weyl-squared term 
\begin{equation} \label{RenVol4}
    \frac{1}{32\pi^2}\int \limits_{M_4} d^4x \sqrt{-g} \,W^{\alpha\beta\mu\nu}W_{\alpha\beta\mu\nu}= \chi(M_4)-\frac{3}{4\pi^2}V_{\rm{ren}}\,
\end{equation}
up to the Euler characteristic of the manifold $\chi(M_4)$.

Thought of as a gravity theory, the left-hand side of eq.\eqref{RenVol4} is proportional to the action of Conformal Gravity, $I_{\rm CG}=\int_{M_{4}} \sqrt{-g}\,W^2$, whose field equations are of higher order in derivatives of the metric tensor.

In point of fact, an arbitrary variation of the previous action produces the equations of motion in the bulk plus a surface term
\begin{equation}
    \delta I_{\rm CG} =\int \limits_{M_4} d^4x \sqrt{-g}\,B^{\mu \nu} \delta g_{\mu \nu} + \int \limits_{\partial M_4}n_{\mu}\Theta^{\mu}\,,
\end{equation}
where $B^{\mu \nu}$ is the Bach tensor, which is a covariantly conserved, symmetric and traceless object. This tensor is conformally covariant under Weyl rescalings of the metric only in four dimensions, and it is defined as
\begin{equation}
    B^{\mu}_{\nu}=S^{\sigma}_{\lambda} W^{\mu\lambda}_{\,\,\,\,\nu \sigma}+\Box S^{\mu}_{\nu}-\nabla^{\sigma}\nabla^{\mu}S_{\sigma\nu},
\end{equation}
with $S_{\mu \nu}$ being the Schouten tensor, which, in terms of the Ricci tensor and the Ricci scalar, is written as
\begin{equation}
    S_{\mu \nu}=\frac{1}{2}\left(R_{\mu \nu}-\frac{1}{6}Rg_{\mu \nu}\right)\,.
\end{equation}

$\Theta^{\mu}$ is a boundary term and $n_{\mu}$ is a normal vector to the boundary.\\
In six dimensions, a formula for Renormalised Volume was proposed by Chang, Qing, and Yang \cite{Chang:2005ska}.

It is straightforward to check that Einstein-AdS spaces
\begin{equation}
    R_{\mu \nu}=-\frac{3}{\ell^2}g_{\mu \nu}\,,
    \label{Einstein-AdS}
\end{equation}
satisfy the field equations of Conformal Gravity, defining a generic sector of the theory. The length $\ell$ is a characteristic scale given by the AdS radius.

From a holographic standpoint, in Einstein gravity, the description of an asymptotically AdS space is given in terms of the Fefferman-Graham coordinate frame~\cite{AST_1985__S131__95_0} 
\begin{equation}
ds^2 = \frac{\ell^2}{z^2} \left(dz^2 + \bar{g}_{ij} \left(z,x\right) dx^i dx^j\right)\,,
\label{FGmetric}
\end{equation}
which endows the boundary with a conformal structure. The boundary metric $\bar{g}_{ij}(z,x)$ is a series in the $z$ holographic coordinate, which starts with the corresponding source $g_{(0)ij}$ and, in the case of  four dimensions, takes the particular form
 \begin{equation}
\bar{g}_{ij}(z,x)=
g_{(0)ij}(x)+\frac{z^{2}}{\ell^2}\,g_{(2)ij}(x)+\frac{z^{3}}{\ell^3}\,g_{(3)ij}(x)+\cdots \,.
\label{FGexpansion}
\end{equation}
The mode at the holographic order $z^3$, i.e., $g_{(3)ij}$, represents the holographic stress tensor, whose vanishing trace is determined by the field equations.
The geometry characterised by the holographic data $g_{(0)ij}$ is assumed to be smooth, i.e., free of higher codimension defects.

In turn, the holographic consistency of the equations of motion in Conformal Gravity requires an additional coefficient $g_{(1)ij}$

\begin{equation}
\bar{g}_{ij}(z,x)=
g_{(0)ij}(x)+\frac{z}{\ell}\,g_{(1)ij}(x)+\frac{z^{2}}{\ell^2}\,g_{(2)ij}(x)+\frac{z^{3}}{\ell^3}\,g_{(3)ij}(x)+\cdots \,,
\label{FGexpansionCG}
\end{equation}
where the holographic stress tensor also contains a response associated with the new source.

It was argued in Ref.\cite{Maldacena:2011mk} that Neumann boundary conditions for the metric $\bar{g}_{ij}$ provide a mechanism to turn Conformal Gravity into Einstein theory. Indeed, the condition $\partial_z\bar{g}_{ij}=0$ implies that $g_{(1)ij}=0$, such that the metric \eqref{FGexpansionCG} reduces to the expansion \eqref{FGexpansion} in Einstein gravity.

At the level of the action, the Einstein condition on the spacetime may be implemented either directly by Eq.\eqref{Einstein-AdS} or by the vanishing of the traceless Ricci tensor
\begin{equation}
    H_{\mu \nu}=R_{\mu \nu}-\frac{1}{4}g_{\mu \nu}R\,.
\end{equation}
The latter equation readily implements the Neumann boundary condition by Maldacena \cite{Anastasiou:2020zwc}. It also implies the rest of the holographic relations for asymptotically AdS spaces in Einstein gravity, including the fact that the holographic stress tensor  $g_{(3)ij}$ is transverse and traceless.

From the general expression for the Weyl tensor 

\begin{equation}
W_{\mu \nu}^{\alpha \beta}=R_{\mu \nu}^{\alpha \beta}-S_{\mu}^{\alpha}\delta
_{\nu}^{\beta}+S_{\nu}^{\alpha}\delta _{\mu}^{\beta}+S_{\mu}^{\beta}\delta _{\nu}^{\alpha}-S_{\nu}^{\beta}\delta
_{\mu}^{\alpha}\,,  \label{W}
\end{equation}
for Einstein-AdS, one obtains a reduced form of this tensor
\begin{equation}
W
_{(E)\mu \nu}^{\alpha \beta}=R_{\mu \nu}^{\alpha \beta}+\frac{1}{\ell^2}{\delta}_{\mu \nu}^{\alpha \beta}\,. \label{W_E}
\end{equation}

From here on, we use the superscript $(E)$ to denote the Einstein sector. Therefore, the Renormalised Volume for Euclidean Einstein-AdS is given as
\begin{equation}
    \frac{1}{32\pi^2}\int \limits_{M_4} d^4x \sqrt{g} \,W_{(E)}^{\alpha\beta\mu\nu}W_{(E)\alpha\beta\mu\nu}= \chi(M_4)-\frac{3}{4\pi^2\ell^{4}}V^{(E)}_{\rm{ren}}\,.
\end{equation}
The above expression is connected to the definition of the MacDowell-Mansouri action \cite{MacDowell:1977jt,Stelle:1979aj}. 
\begin{equation*}
I_{\text{MM}}=-\frac{\ell ^{2}}{256\pi G}\int\limits_{M_4}d^{4}x\sqrt{g}%
\,\delta _{ \mu _{1}\cdots \mu _{4} }^{ \nu _{1}\cdots
\nu _{4} }\left( R_{\nu _{1}\nu _{2}}^{\mu _{1}\mu _{2}}+\frac{\delta
_{\nu _{1}\nu _{2}}^{\mu _{1}\mu _{2}}}{\ell ^{2}}\right) \left( R_{\nu
_{3}\nu _{4}}^{\mu _{3}\mu _{4}}+\frac{\delta _{\nu _{3}\nu _{4}}^{\mu
_{3}\mu _{4}}}{\ell ^{2}}\right) \,.
\end{equation*}

Using the identity \footnote{Here, the antisymmetric generalisation of the Kronecker delta is defined as $\delta^{\nu_1\dots\nu_k}_{\mu_1\dots\mu_k}\equiv\det\left[\delta^{\nu_1}_{\mu_1}\cdots\delta^{\nu_k}_{\mu_{k}}\right].$}
\begin{equation}
 \delta_{\nu_{1}\nu_{2}\nu_{3}\nu_{4}}^{\mu_{1}\mu_{2}\mu_{3}\mu_{4}} W^{
\nu_{1}\nu_{2}}_{(E)\mu_{1}\mu_{2}}W^{\nu_{3}\nu_{4}}_{(E)\mu_{3}\mu_{4}} = \frac{16}{\ell^{2}}R + \frac{96}{\ell^{4}}+4 \mathcal{G}\,,
\end{equation}
with 
\begin{equation} \label{GB}
\mathcal{G}= R^2 - 4 R_{\mu \nu}R^{\mu \nu}+ R^{\mu \nu \alpha \beta}R_{\mu \nu \alpha \beta} \,,
\end{equation}
the MacDowell-Mansouri action becomes the Einstein-AdS gravity action augmented by the Gauss-Bonnet invariant with a fixed coupling,
\begin{equation}\label{eq:topologicalaction}
I_{MM}=I_{\rm{top}}=-\frac{1}{16 \pi G}\int\limits_{ M_4} d^4 x \sqrt{g} \left(R+\frac{6}{\ell^{2}} + \frac{\ell^{2}}{4} \mathcal{G}\right)\,, 
\end{equation}
In the Einstein branch, $R=-12/\ell^{2}$, and $\mathcal{G}=W^{2}_{(E)}+24/\ell^{4}$, so the constant terms cancel leading to 
\begin{equation}\label{eq:top-action-einstein}
    I_{\rm{top}}= \frac{\ell^{2}}{64\pi G}\int\limits_{M_4}d^{4}x\sqrt{g}\, W^{2}_{(E)}\,.
\end{equation}

The addition of the Gauss--Bonnet term does not affect the equations of motion. The coupling $\ell^{2}/4$ is singled out by a finite action for asymptotically AdS Einstein spaces \cite{Aros:1999id}. In this way, the consistent inclusion of a topological term of the Euler class renormalises the gravitational action in a fully covariant form. This defines the concept of Topological Renormalisation \cite{Olea:2005gb,Miskovic:2009bm}. In point of fact, the Fefferman-Graham coordinate frame \eqref{FGmetric}-\eqref{FGexpansion}, when plugged into the on-shell variation of the MacDowell-Mansouri action
into\begin{equation}
\delta I_{\rm{top}} = \frac{\ell^2}{64\pi G} \int\limits_{M_4} d^3x \sqrt{-h}\, n_{\nu_1} \delta _{ \mu _{1}\cdots \mu _{4} }^{ \nu _{1}\cdots
\nu _{4} } \delta g^{\mu_2 \lambda} \Gamma^{\mu_1}_{\lambda \nu_2} \left( R^{\mu_3 \mu_4}_{\nu_3\nu_4}+\frac{1}{\ell^2}\delta^{\mu_3 \mu_4}_{\nu_3\nu_4}\right) \,,
\end{equation}
reproduces the holographic content of four-dimensional AdS gravity, that is,
\begin{equation}
    \delta I_{\rm{top}} = \frac{1}{2} \int\limits_{\partial M_4} d^3 x \sqrt{-g_{(0)}}\left(\frac{3}{16\pi G\ell} g_{(3)}^{i j}\delta g_{(0)ij}\right) \,.
\end{equation}
In other words, the total variation of the action is free from infrared divergences and duly expressed in terms of the holographic stress tensor $T^{ij}=3 g^{ij}_{(3)}/(16\pi G\ell)$ and the variation of the holographic source $g_{(0)ij}$ \cite{Balasubramanian:1999re,deHaro:2000vlm}. 

A finite term proportional to the Euler characteristic drops out of this variation at fixed topology and couplings. This is more evident when comparing thermodynamic quantities. For smooth manifolds, the topological action produces a constant shift in the entropy, as illustrated in \cite{Olea:2005gb} for the Kerr--AdS black hole. In the presence of conical deficits, the finite difference remains after applying the Euler normalisation appropriate to smooth fillings \cite{Hale:2025xnh}. The slowly accelerating AdS C-metric provides a setting in which to trace this residual difference to the curvature supported on the conical axes. Unequal string tensions drive the acceleration, while the absence of an acceleration horizon allows the Euclidean time period to be fixed by regularity at the black hole horizon. Even in this regime, the thermodynamic description requires a choice of time normalisation and a specification of which deficit variations are allowed. The first law was formulated with fixed deficits \cite{Appels:2016uha}, and later extended to include varying string tensions and their role as conjugate thermodynamic lengths \cite{Appels:2017xoe}. See also \cite{Ball:2020vzo} for the analysis of the charged C-metric first law using boost time. 
Holographic calculations related the mass and free energy for a suitable time normalisation \cite{Anabalon:2018ydc,Anabalon:2018qfv,Cassani:2021dwa}. In covariant phase space, Kim et al. \cite{Kim:2023ncn} showed how the usual time normalisation is recovered at fixed total deficit and how holographic counterterms induce a corner contribution to the charges where the strings meet conformal infinity. 

The role of the defect is more evident in three dimensions. The accelerating BTZ black hole \cite{Astorino:2011mw, Arenas-Henriquez:2022www} illustrates how the defect can enter the renormalisation of the Euclidean action itself. The domain wall--the codimension-one analogue to the cosmic string--reaches the conformal boundary, and the usual asymptotic counterterms leave a subleading divergence. The wall contribution, including the internal Gibbons--Hawking--York terms, cancels this divergence and takes the Nambu--Goto form on shell by the Israel junction conditions \cite{Arenas-Henriquez:2023hur, Cisterna:2023qhh}. In the generalised Fefferman--Graham gauge, the divergence is cancelled by a corner counterterm where the wall meets the conformal boundary \cite{Arenas-Henriquez:2024ypo}. Both calculations recover the quantum statistical relation. 

These results motivate a closer look at how the two renormalisation prescriptions account for the conical axes in four dimensions. Hale et al. \cite{Hale:2025xnh} found a finite difference between the holographically and topologically renormalised Euclidean actions of charged accelerating black holes after applying Euler normalisation for smooth filling. We examine this mismatch for the neutral AdS C-metric. By separating the regular curvature from the contributions on the axes, we relate the residual difference to the tension-weighted Euler terms of the renormalised worldsheet areas and identify it with the Gauss--Bonnet entropy contribution at the polar conical points.

This paper is organised as follows: Section \ref{sec:2} deals with black hole thermodynamics of regular horizons by directly comparing the Gauss-Bonnet term with holographic counterterms. Section \ref{sec:3} introduces the C-metric solution to the Einstein equations, and highlights its main features. It also shows the results for the different prescriptions of Euclidean action evaluated in the C-metric. Section \ref{sec:4} defines the notion of Renormalised Area, as a consequence of the addition of a topological term in the bulk. Section \ref{sec:5} studies the contribution from conical defects to the black hole thermodynamics. The last section is devoted to some final remarks.

\section{Counterterms in AdS gravity and Regular Thermodynamics} \label{sec:2}
Within a more physical framework, the relation to a renormalised action based on the addition of counterterms \cite{Balasubramanian:1999re,Emparan:1999pm} is possible only if one breaks general covariance of the spacetime by introducing a boundary. This is explicitly carried out, e.g., in Gaussian coordinates
\begin{equation}
    ds^2=N^2(z)dz^2+h_{ij}(x,z)dx^i dx^j \,.
\end{equation}
In asymptotically AdS spaces, the lapse function $N$ is given by $N=\ell/z$, while the induced metric takes the form $h_{ij}(x,z)=\ell^2\bar{g}_{ij}/z^2$.

This comes from insisting on a Dirichlet problem for the boundary metric $h_{ij}$ which, in principle, is ill-defined for asymptotically AdS spacetimes \cite{Papadimitriou:2005ii}. The inclusion of the Gibbons-Hawking term plus boundary counterterms, --only when properly resummed-- can be expressed as bulk-covariant quantities. The explicit proof goes as follows: consider a smooth Euclidean four-manifold with boundary. In our conventions, the Chern-Gauss-Bonnet theorem takes the following form
\begin{equation}
    \int\limits_{M_4} d^{4}x \sqrt{g}\,\mathcal{G}=32\pi^2 \chi(M_4)-\int\limits_{\partial M_4}B_{3}\,.
\end{equation}
where $B_3$ denotes the density of the boundary Chern form,
\begin{equation}
B_{3} =4\sqrt{h} \delta _{ j_{1}j_{2}
j_{3}}^{i_{1}i_{2} i_{3}}K_{i_{1}}^{j_{1}}%
\left( \frac{1}{2}\mathcal{R}%
_{i_{2}i_{3}}^{j_{2}j_{3}}-\frac{1}{3}K_{i_{2}}^{j_{2}}K_{i_{3}}^{j_{3}}\right)\,.
\end{equation}
The Chern density is a given polynomial of the extrinsic curvature, defined by
\begin{equation}    
K_{ij}=h_{i}^{\mu}h_{j}^{\nu}\nabla_{\mu}n_\nu \,,
\end{equation}
where $n_{\mu}=-N\delta^{z}_{\mu}$ is the outward-pointing unit normal to the boundary. The intrinsic Riemann tensor is computed from the boundary metric, $\mathcal{R}_{kl}^{ij}=\mathcal{R}_{kl}^{ij}(h)$.

The Einstein-AdS action, renormalised with a topological term, can be equivalently written as
\begin{equation}\label{eq:kounterterm}
  I_{\rm{top}}+\frac{\pi \ell^{2}}{2G}\chi(M_{4})= I_{\rm{EH}}+\frac{ \ell^{2}}{64\pi G}\int\limits_{\partial M_4} d^3x\,B_{3}\,.
\end{equation}
Adding and subtracting the Euclidean Gibbons-Hawking term, the above action takes the form
\begin{equation}\label{eq:CTs}
  I_{\rm{top}}+\frac{\pi \ell^{2}}{2G}\chi(M_{4})= I_{\rm{EH}}-\frac{1}{8\pi G}\int\limits_{\partial M_4} d^3x \sqrt{-h}K +\int\limits_{\partial M_4} d^3x\,\mathcal{L}_{\rm{ct}}\,,
\end{equation}
where $\mathcal{L}_{\rm{ct}}$ is the resummation of local counterterms in terms of the extrinsic curvature
\begin{equation}
\mathcal{L}_{\rm{ct}} =\frac{\ell^2}{16\pi G}\sqrt{-h} \delta _{ j_{1}j_{2}
j_{3}}^{i_{1}i_{2} i_{3}}K_{i_{1}}^{j_{1}}%
\left( \frac{1}{2}\mathcal{R}%
_{i_{2}i_{3}}^{j_{2}j_{3}}-\frac{1}{3}K_{i_{2}}^{j_{2}}K_{i_{3}}^{j_{3}}+\frac{1}{\ell^2}\delta_{i_{2}}^{j_{2}}\delta_{i_{3}}^{j_{3}}\right)\,.
\end{equation}
This becomes manifest once one considers the asymptotic expansion of $K^i_j$ as
\begin{equation}
K_{j}^{i}=\frac{1}{\ell }\,\delta _{j}^{i}+\ell \, \mathcal{S}_{j}^{i}(h) +\mathcal{O}(z^3)\,,
\label{K.expanded}
\end{equation}
where $\mathcal{S}_{j}^{i}(h)$ is the Schouten tensor of the boundary metric.

In point of fact, by expanding and further collecting the first two leading orders in the boundary curvature, this expression results in ~\cite{Balasubramanian:1999re, Emparan:1999pm}

\begin{equation}
\mathcal{L}_{\rm{ct}}=\frac{\sqrt{h}}{8\pi G }\left(\frac{2}{\ell}+
\frac{\ell}{2} \mathcal{R}
 \right)\,+\cdots\,,
\end{equation}
which are the standard regulating counterterms in AdS$_{4}$ gravity, such that the r.h.s. of Eq.\eqref{eq:CTs} is the renormalised AdS action $I_{\rm{ren}}$ in the context of gauge/gravity duality.

Hence, on a smooth manifold $I_{\rm{top}}+\frac{\pi \ell^{2}}{2G}\chi(M_4)= I_{\rm{ren}}$, with
\begin{equation}\label{eq:Iren}
  I_{\rm{ren}}= I_{\rm{EH}}-\frac{1}{8\pi G}\int\limits_{\partial M_4} d^3x \sqrt{h}K +\frac{1}{8\pi G }\int\limits_{\partial M_4} d^3x\,\sqrt{h}\left(\frac{2}{\ell}+
\frac{\ell}{2} \mathcal{R}
 \right)\,.
\end{equation}
For an Einstein manifold, the on-shell bulk density is $-(R+6/\ell^{2})=6/\ell^{2}$. The above expression allows us to express the renormalised action in terms of the Renormalised Volume
\begin{eqnarray}
   I_{\rm{ren}} &= & \frac{3}{8 \pi G \ell^{2}}V^{(E)}_{\rm{ren}} \\
                &= &I_{\rm{top}} + \frac{\pi \ell^{2}}{2G }\chi(M_{4})  \,.
\end{eqnarray}
As it is clear from this expression, on a smooth filling, holographically and topologically renormalised actions differ by the Euler term. At fixed topology, $\ell$ and $G$, this gives a constant shift to Wald entropy. More generally, conical defects may extend through the bulk to the conformal boundary. Each action then receives contributions supported on the codimension-two surfaces, which must be separated from the regular contribution \cite{Fursaev:2013fta}. For the AdS C-metric, we will show how this decomposition relates the finite action mismatch to the renormalised areas of the string worldsheets and their Euler characteristics weighted by the conical deficits. We refer to the inclusion of such contributions as \textit{singular}. These singular contributions vary with the string tensions, even when the topology remains fixed.

\section{The \texorpdfstring{AdS$_{4}$}{AdS4} C-metric} \label{sec:3}

The AdS C-metric belongs to the Pleba\'nski--Demia\'nski family of solutions and describes a pair of uniformly accelerating black holes in an asymptotically AdS spacetime \cite{Plebanski:1976gy,Dias:2002mi,Podolsky:2006px,Griffiths:2005qp,Krtous:2005ej}. The acceleration is driven by conical deficits along the polar axis, which may be interpreted as cosmic strings or struts.  We restrict our attention to the neutral, non-rotating solution, parametrised by the mass parameter $m$, the acceleration $A$, the AdS radius $\ell$, and the conical parameter $\mathcal{K}$. In Schwarzschild-like coordinates, the line element is \cite{Hong:2003gx,Appels:2016uha}
\begin{equation}
ds^{2}=\frac{1}{\Omega^{2}}\left[ -\frac{f(r)}{\alpha^{2}}dt^{2}+\frac{1}{f(r)}dr^{2}+ r^{2}\left( \frac{d\theta^{2}}{g(\theta)}+g(\theta)\sin^{2}\theta\frac{d\phi^{2}}{\mathcal{K}^{2}}\right)\right], \label{eq:accbh4d}
\end{equation}
where 
\begin{eqnarray}
\Omega & = & 1+Ar\cos{\theta}, \nonumber \\ 
f(r) &=& (1-A^{2}r^{2})\left(1-\frac{2m}{r}\right)+\frac{r^{2}}{\ell^{2}}, \nonumber \\
g(\theta)&=& 1+2mA\cos{\theta}, \quad \alpha=\sqrt{1-A^{2}\ell^{2}} \,.
\end{eqnarray}
The azimuthal coordinate has periodicity $\phi \sim \phi +2\pi$. The parameters $m$ and $A$ determine the local geometry, while $\mathcal{K}$ fixes the global identification of the azimuthal coordinate, and hence, the conical deficits. The conformal factor $\Omega$ diverges at $r=-1/A\cos{\theta}$ where the conformal boundary is located. The factor $\alpha$ in \eqref{eq:accbh4d} fixes the normalisation of the timelike Killing vector at conformal infinity \cite{Appels:2016uha,Anabalon:2018ydc} and can be obtained by mapping the $m=0$ solution to global AdS. 
 
 In principle, there are two horizons depending on the interplay between mass, acceleration, and cosmological constant. We restrict ourselves to the \textit{slow acceleration} approximation where $A\ell <1$, in which the exterior region of the black hole contains no acceleration horizon. In this case, the black hole horizon equation determines the mass parameter as 
 \begin{equation}\label{eq:massparameter}
     m = \frac{r_{+}}{2}\Bigg[1+\frac{r_{+}^{2}}{\ell^{2}(1- A^{2}r^{2}_{+})}\Bigg] \,.
 \end{equation}
After a Wick rotation, regularity at the Euclidean horizon fixes the inverse temperature,
\begin{equation}\label{eq:invtemp}
   \beta = \frac{4 \pi \alpha}{f'(r_{+})}\,. 
\end{equation}
The physical mass may be computed either from the Ashtekar-Magnon-Das formula or from the holographic stress tensor, see for instance \cite{Ashtekar_1984,Ashtekar:1999jx,Jatkar:2014npa,Kim:2023ncn,Hale:2025xnh}. With the asymptotic time normalization used in \eqref{eq:accbh4d}, one obtains 
\begin{equation} \label{eq:mass}
M=\frac{\alpha m}{G \mathcal{K}}.    
\end{equation}
The Schwarzschild-AdS result is recovered in the limit $A\to 0$ and $\mathcal{K}\to 1$.
The parameter $\mathcal{K}$ controls the conical geometry of the polar axes. At the north and south poles,
\begin{equation} \label{eq:epoles}
    \theta_{+}=0 \,, \quad     \theta_{-}=\pi \,,
\end{equation}
the angular function takes the values 
\begin{equation}
    g_{\pm}:=g(\theta_{\pm})=1\pm 2m A\,.
\end{equation}
The ratio between the proper circumference and proper radius near each pole is $2\pi g_{\pm}/\mathcal{K}$. It is therefore useful to introduce the two cone-opening parameters
\begin{equation}\label{eq:coneopenning}
    a_{\pm}= \frac{g_{\pm}}{\mathcal{K}}=\frac{1\pm 2m A}{\mathcal{K}}\,.
\end{equation}
The corresponding deficit angles and cosmic string tensions are 
\begin{equation}\label{eq:deltamu}
    \delta_{\pm}=2\pi (1-a_{\pm}), \qquad \mu_{\pm}=\frac{\delta_{\pm}}{8\pi G}=\frac{1}{4 G}\left(1-\frac{g(\theta_{\pm})}{\mathcal{K}}\right) =\frac{1}{4}\left(1-\frac{1\pm 2mA}{\mathcal{K}}\right) \,.
\end{equation}
It is also useful to introduce the sum and difference of tensions
\begin{equation}\label{eq:musumdiff}
    \mu _{+}+ \mu_{-}=\frac{1}{2 G}\left(1-\frac{1}{\mathcal{K}}\right)\,,\quad 
      \mu _{+}- \mu_{-}=\frac{mA}{G \mathcal{K}}\,.
\end{equation}
Thus, $\mathcal{K}$ fixes the total conical deficit, while $m A$ measures the imbalance between the two tensions that produce the acceleration. One may remove either conical singularity by an appropriate choice of $\mathcal{K}$. For example,
\begin{equation}
    \mathcal{K}=g_{+}=1+2m A\,,
\end{equation}
making the north axis regular and leaving a cosmic string on the south axis with tension 
\begin{equation}
    \mu_{-}=\frac{m A}{G\mathcal{K}}\,.
\end{equation}
In the present work, we shall not impose this regularity condition. Both axes will instead be retained as codimension-two conical loci, with opening parameters $a_{+}$ and $a_{-}$.

At fixed $t$ and $r$, the angular part of the metric is topologically a two-sphere with conical singularities at its north and south poles. For generic values of $a_{\pm}$, it should therefore be regarded as a two-dimensional cone-manifold, or cone-sphere. For the special choice
\begin{equation}\label{eq:quantisation}
    a_{\pm}=\frac{1}{n_{\pm}}\, \quad n_{\pm}\in \mathbb{N}\,,
\end{equation}
the two cone angles are orbifold angles, and the surface becomes the weighted projective space $\Sigma=\mathbb{WCP}^{1}_{[n_{-},n_{+}]}$ \cite{Cassani:2021dwa}. The generic C-metric considered here is not assumed to obey the quantisation condition \eqref{eq:quantisation}; the spindle provides a useful orbifold specialisation to more general cone-manifolds. We will come back to this point later.

\subsection{On-shell Euclidean Actions}
We now review the two different prescriptions that render the Euclidean action finite. The first is the standard holographic renormalisation procedure based on intrinsic counterterms at the conformal boundary. The second is the topological prescription, in which the Einstein-AdS action is supplemented by the four-dimensional Gauss-Bonnet invariant. 
\subsubsection{Holographic renormalisation}

For the C-metric \eqref{eq:accbh4d},  the conformal boundary is not located at constant $r$. Therefore, it is convenient to perform a suitable change of coordinates to facilitate the evaluation of Euclidean action from the horizon to the conformal boundary. 

We choose the transformation
\begin{equation}
    x=\cos\theta\,,\quad\frac{1}{\rho}=\frac{1}{r}+Ax\,, \quad\text{such that}\quad r=\frac{\rho}{1-A\rho x}\,.
\end{equation}
Using the above, the line element \eqref{eq:accbh4d} in Euclidean signature becomes
\begin{equation}\label{eq:c-metric-yx}
    ds_{E}^{2}=\frac{d\rho^2}{H(\rho,x)}+\frac{2A\rho^2}{H(\rho,x)}d\rho dx+\frac{H(\rho,x)}{\alpha^2}d\tau^2+\rho^2\left(\left(\frac{A^2\rho^2}{H(\rho,x)}+\frac{1}{P(x)}\right)dx^2+\frac{P(x)}{\mathcal{K}^2}d\phi^2\right)\,,
\end{equation}
where 
\begin{equation}
    H(\rho,x)=\rho^2\left[\left(\frac{1}{\rho^2}-\frac{2Ax}{\rho}-\left(1-x^2\right)A^2\right)\left(1-2m\left(\frac{1}{\rho}-Ax\right)\right)+\frac{1}{\ell^2}\right]\,,
\end{equation}
and
\begin{equation}
    P(x)=(1-x^{2})(1+2m Ax)\,.
\end{equation}

In these coordinates $-1\le x\le 1$ and $\rho_h(x)\le\rho\le\infty$, where $\rho_h(x)$ is evaluated at the horizon as 
\begin{equation}
    \rho_h(x)=\frac{r_+}{1+Ar_+x}\,,
\end{equation} 
while $0\le\tau\le\beta$ and $0\le\phi\le 2\pi$. These coordinates cover the complete axial disks. In particular, the north axis continues through $1/r=0$ before reaching the conformal infinity, see for instance \cite{Hubeny:2009kz} or, in the case of three dimensions, \cite{Arenas-Henriquez:2023hur}. 

Evaluating the Einstein action, the Gibbons-Hawking term, and the two intrinsic contributions of \eqref{eq:Iren}, we get \cite{Anabalon:2018ydc,Cassani:2021dwa} 
\begin{equation}
I^{E}_{\rm{ren}} = \frac{\beta}{2 \alpha G \mathcal{K}}\left( m(1- 2 A^{2} \ell^{2}) - \frac{r^3_{+}}{\ell^{2}(1-A^{2}r_{+}^2)^2}\right)\,. \label{eq:euclidean-ct}
\end{equation}
Here $\beta$ is the period of the normalised Euclidean time \eqref{eq:invtemp}. The standard volume and curvature counterterms suffice for this on-shell evaluation. 

\subsection{Topological renormalisation}

The C-metric can be directly evaluated into the bulk action \eqref{eq:topologicalaction}, without the need for the FG expansion for the metric, as it is a fully covariant expression. On a regular Einstein geometry, we have $R=-12/\ell^{2}$ and $\mathcal{G}=W^{2}+24/\ell^{2}$. For the C-metric, $W^{2}=48m^{2}/\rho^6$, and $\sqrt{g}=\rho^2/(\alpha\mathcal{K})$. Evaluating the integral in \eqref{eq:top-action-einstein}, we obtain 
\begin{equation}
 I^{E}_{\rm{top}}=\frac{-\beta m^{2}\ell^{2} }{\alpha \mathcal{K}G} \left(\frac{1}{r_{+}^3}+\frac{A^{2}}{r_{+}}\right)\,. \label{Euc-GB}
\end{equation}

\subsection{Comparison of the two actions}
Now, let us compare the results of both renormalisation prescriptions explicitly. Let us define
\begin{equation}
   D_{+}:= 1- A^{2}r^{2}_{+}\,. 
\end{equation}
The horizon equation and the radial derivative of the metric function at the horizon give
\begin{equation}
 m=\frac{r_{+}}{2}\left( 1+\frac{r^{2}_{+}}{\ell^{2}D_{+}}\right)\,, \quad f'(r_{+}) =2m \left( \frac{1}{r^{2}_{+}}+A^{2}\right) +2 r_{+} \left(\frac{1}{\ell^{2}} -A^{2}\right) \,.
\end{equation}
Using the above expression and the regularity of the Euclidean time, we can see that the actions \eqref{eq:topologicalaction} and \eqref{eq:Iren} differ by a constant
\begin{equation}
I_{\rm{top}}-I_{\rm{ren}}=-\frac{\beta \ell^{2}f'(r_{+})}{4G \alpha \mathcal{K}}=-\frac{\pi \ell^{2}}{G \mathcal{K}}    \,.
\end{equation}
This is the mismatch emphasised in \cite{Hale:2025xnh}, where the remaining term is a finite Euler contribution. The conformally compactified Euclidean black hole exterior has the underlying topology
\begin{equation}
    M_4\simeq D^{2}\times S^{2}, \quad \chi(M_4)=2\,.
\end{equation}
It is useful to normalise the topological action as in the smooth Schwarzschild-AdS case, that is
\begin{equation}
    I_{T}:=I_{\rm{top}}+\frac{\pi \ell^{2}}{2G}\chi(M_4)=I_{\rm{top}}+\frac{\pi \ell^{2}}{G}\,.
\end{equation}
Then, the difference between the actions is
\begin{equation}\label{eq:mismatch}
    I_{T}-I_{\rm{ren}}=\frac{\pi \ell^{2}}{G}\left( 1-\frac{1}{\mathcal{K}}\right)=2\pi\ell^{2}\mu, \qquad \mu:=\mu_{+}+\mu_{-}\,.
\end{equation}
It is easy to see that the distinction persists even when $A \to 0$ with $\mathcal{K}>1$. Then, there is an equal tension at the two poles. The azimuthal integral of every regular curvature density is multiplied by $1/\mathcal{K}$, while the ordinary Euler characteristic remains two. 

\section{Renormalised Area}\label{sec:4}

A conical defect separates a curvature integral into a regular part and a contribution supported on a codimension-two surface $\Sigma$. The surface carries the induced metric $\gamma_{ab}$, with $\gamma=\det (\gamma_{ab})$. We denote the opening fraction as  $a$, so that the cone angle is $2\pi a$ and the deficit angle is $2\pi (1-a)$. The corresponding physical tension is $\mu =(1-a)/4G$, as discussed in \autoref{sec:3}. 
To regulate the noncompact integrals, we introduce a common outer boundary at $\rho=\rho_{c}$. Let $M^{(a)}_{4,c}$ denote the truncated conical geometry and set
\begin{equation}
    \Sigma_{c}=\Sigma \cap M^{(a)}_{4,c}\,, \qquad C_{c}=\Sigma \cap \{\rho=\rho_{c}\}=\partial \Sigma_{c}\,.
\end{equation}
For the axial disks below, the Euclidean time circle closes smoothly at the horizon at the Hawking period, so $C_{c}$ is their only boundary. We write $M_{\rm{reg},c}$ for the regular part of the same conical geometry, retaining its angular range. A smooth regulator of the cone is introduced at fixed $\rho_{c}$, and the conical limit is taken before removing the outcutoff. At finite $\rho_{c}$, the scalar curvature satisfies
\begin{equation}\label{eq:ricci-decomposed}
\int\limits_{M_{4,c}^a} \sqrt{g}R^{(a)}=\int\limits_{M_{\rm{reg},c}}d^{4}x \sqrt{g}R + 4\pi (1-a)  \mathcal{A} \left[\Sigma_{c}\right]\,, \quad 
\end{equation}
where the singular contribution is given by the area functional of a codimension-two surface $\Sigma_{c}$
\begin{equation}
\mathcal{A} \left[\Sigma_{c}\right]=\int\limits_{\Sigma_{c}} d^{2}y \sqrt{\gamma}\,.
\end{equation}
 Let $e_{a}^{\mu}=\partial X^{\mu}/\partial y^{a}$ be tangent vectors to $\Sigma$, so that $\gamma_{ab}=g_{\mu \nu}e_{a}^{\mu} e_{b}^{\nu}$. For orthonormal normal vectors $n_{A}^{\mu}$, satisfiying $g_{\mu \nu}n^{\mu}_{A} n^{\nu}_{B}=\delta_{AB}$, and $g_{\mu \nu} n^{\mu}_{A} e^{\nu}_{a}=0$. The extrinsic curvature and their traces are
\begin{equation}
 k^{(A)}_{ab}=e^{\mu}_{a}e_{b}^{\nu}\nabla_{\mu}n^{(A)}_{\nu}\,, \qquad k^{(A)}=\gamma^{ab}k^{(A)}_{ab}\,.
\end{equation}
These quantities are evaluated in a smooth regularisation of the defect, with $\nabla$ the ambient Levi-Civita connection. Tangential indices $\{a,b\}$ are raised with $\gamma^{ab}$, while normal indices $\{A,B\}$ with $\delta^{AB}$. We can evaluate the quadratic invariants up to linear terms in the deficit \cite{Fursaev:2013fta}
\begin{align}
\int\limits_{M_{4,c}^a} d^{4}x \sqrt{g}
\left(Rie^{(a)}\right)^{2} &=\int\limits_{M_{\rm{reg},c}} d^{4}x\sqrt{g}\, Rie^{2}  + 8\pi \left(1-a\right) \int\limits_{\Sigma} d^{2}y \sqrt{\gamma} \left(R^{AB}_{AB} - k^{\left(A\right)}_{ab} k^{ab}_{\left(A\right)}\right) + ... \,, \\
\int\limits_{M_{4,c}^a} d^{4}x \sqrt{g} \left(Ric^{(a)}\right)^{2}&= \int\limits_{M_{\rm{reg},c}} d^{4}x \sqrt{g}\,  Ric^{2} + 4\pi \left(1-a\right) \int\limits_{\Sigma} d^{2}y \sqrt{\gamma} \left(R_{AA} -\frac{1}{2} k^{\left(A\right)} k_{\left(A\right)}\right) +...\,. \nonumber \\
\int\limits_{M_{4,c}^a} d^{4}x \sqrt{g} \left(R^{(a)}\right)^{2} &= \int\limits_{M_{\rm{reg},c}} d^{4}x \sqrt{g}\, R^{2} + 8\pi \left(1-a\right) \int\limits_{\Sigma}\sqrt{\gamma}\,R +...\,.\nonumber
\end{align}
The contractions $R_{AA}=R_{\mu \nu}n^{\mu}_{A}n^{\nu}_{A}$ and $R_{ABAB}=R_{\mu \nu \rho\sigma}n^{\mu}_{A}n^{\nu}_{B}n^{\rho}_{A}n^{\sigma}_{B}$ use the regular ambient curvature. The dots denote regulator-dependent terms of order $(1-a)^{2}$ and higher. The Gauss relation is
\begin{eqnarray}
    R = \mathscr{R}+2 R_{AA} -R_{ABAB} - k^{(A)}k_{(A)} + k^{(A)}_{ab}k^{ab}_{(A)}\,,
\end{eqnarray}
where we denoted intrinsic scalar curvature of $\gamma_{ab}$ by $\mathscr{R}$. Consequently, their Gauss-Bonnet combination has the intrinsic defect term 
\begin{equation}\label{eq:gaussbonnet-decomposed}
\int\limits_{M_{4,c}^a} d^{4}x\, \sqrt{g}\, \mathcal{G}^{(a)} = \int\limits_{M_{\rm{reg},c}} d^{4}x\, \sqrt{g}\, \mathcal{G} + 8 \pi \left(1-a\right) \int\limits_{\Sigma} d^{2}y\,\sqrt{\gamma} \mathscr{R}\,.
\end{equation}
We now insert these relations into the action. The deficit fraction combines with Newton's constant into the physical tension $\mu$, and combining \eqref{eq:ricci-decomposed} with \eqref{eq:gaussbonnet-decomposed}, the topological action, at finite cutoff, gives
\begin{equation}\label{eq:regulatedaction}
    I_{\rm{top}}[M^{(a)}_{4,c}]=I^{\rm{reg}}_{\rm{top}}- \mu\left(\mathcal{A}[\Sigma_{c}]+\frac{\ell^{2}}{2}\int\limits_{\Sigma_{c}}\sqrt{\gamma}\mathscr{R}\right)\,.
\end{equation}
Here $I^{\rm{reg}}_{\rm{top}}(\rho_{c})$ is the action evaluated on the regular curvature densities at the cutoff surface. The Einstein contribution is $-\mu\mathcal{A}[\Sigma_{c}]  $, while the Gauss-Bonnet coupling supplies the intrinsic curvature term. 
To identify the finite area in \eqref{eq:regulatedaction}, we apply the two-dimensional Gauss-Bonnet theorem to the compact surface $\Sigma_{c}$, that is
\begin{equation}
    \int\limits_{\Sigma_{c}}\sqrt{\gamma}\mathscr{R}+2\int\limits_{C_{c}}k_{g}dl=4\pi \chi(\Sigma_{c})=4\pi \chi(\Sigma)\,.
\end{equation}
The boundary geodesic curvature $k_{g}$ is defined with the sign that is positive on a flat disk. Multiplying by $\ell^{2}$, adding $\mathcal{A}[\Sigma_{c}]$ on both sides, and rearranging the identity leads to
\begin{equation}
    \mathcal{A}[\Sigma_{c}]+\frac{\ell^{2}}{2}\int\limits_{\Sigma_{c}}\sqrt{\gamma}\mathscr{R} = \mathcal{A}[\Sigma_{c}]-\ell^{2}\int\limits_{C_{c}}k_{g}dl+2\pi \ell^{2} \chi(\Sigma)\,.
\end{equation}
The cutoff is removed by taking $\rho_{c}\to \infty$, so that $C_{c}$ approaches the conformal boundary and $\Sigma_{c}$ exhaust the complete worldsheet. The area and intrinsic curvature integrals diverge separately, but their combination has the finite limit
\begin{align}
 \lim_{\rho_{c}\to \infty}\left(\mathcal{A}[\Sigma_{c}]-\ell^{2}\int\limits_{C_{c}}k_{g}dl\right)= \mathcal{A}_{\rm{ren}}[\Sigma]+ 2\pi \ell^{2}\chi(\Sigma)\,.
\end{align}
Subtracting this finite Euler term gives the renormalised area formula  \cite{Alexakis:2010zz,Anastasiou:2018mfk,Anastasiou:2022ljq}
\begin{equation}\label{eq:Aren}
    \mathcal{A}_{\rm{ren}}\left[\Sigma\right]=\frac{\ell^{2}}{4}\int\limits_{\Sigma}d^{2}y\, \delta^{ab}_{cd}\left(\mathscr{R}^{cd}_{ab}[\gamma]+\frac{1}{\ell^{2}}\delta_{ab}^{cd}\right)-2\pi \ell^2 \chi(\Sigma)\,.
\end{equation}
This is the codimension-two counterpart of the curvature combination that renormalises the bulk volume. The cancellation removes the area divergence, while the Euler term fixes the finite normalisation. Thus, after removing the cutoff, the defect action \eqref{eq:regulatedaction} becomes
\begin{equation}
    I_{\rm{top}}[M^{(\mu)}_{4}]=I^{\rm{reg}}_{\rm{top}}- \mu (  \mathcal{A}_{\rm{ren}}\left[\Sigma\right]+2\pi \ell^{2}\chi(\Sigma))\,.
\end{equation}
The above is valid for any manifold with conical deficits. 

\subsection{Renormalised worldsheet areas in the AdS C-metric}

We now compute \eqref{eq:Aren} for the C-metric under consideration.  First we separate the two poles of the C-metric, writing $s=+1$ at the north axis and $s=-1$ at the south axis. Then, the worldsheet $\Sigma_{s}$ has induced metric $\gamma^{s}_{ij}$ which corresponds to the $\rho-\tau$ plane, 
\begin{equation}\label{cod2_metric}
    ds^2_{\Sigma_{s}}=\frac{H_{s}(\rho)}{\alpha^2}d\tau^2+\frac{d\rho^2}{H_{s}(\rho)}\,,
\end{equation}
with
\begin{equation}
    H_{s}(\rho)=\frac{\rho^2}{\ell^2}- 2As\left(1+ 2mAs\right)\rho+1+ 6smA-\frac{2m}{\rho}\,.
\end{equation}
The induced metric has a horizon at 
\begin{equation}
    \rho_{h,s}=\frac{r_{+}}{1+s A r_{+}}\,,
\end{equation}
and the truncated disk is $\Sigma_{s,c}=\{\rho_{h,s}\leq\rho\leq\rho_{c}, 0\leq \tau \leq\beta$, with boundary $C_{s,c}$ at $\rho_{c}$. Since $\sqrt{\gamma}=1/\alpha$, its regulated area and intrinsic curvature are
\begin{equation}\label{Area_divergence}
     \mathcal{A}[\Sigma_{s}]=\frac{\beta}{\alpha}(\rho_c-\rho_{h,s})\,,\quad \mathscr{R}_{s}= - H''_{s}= -\frac{2}{\ell^{2}}+\frac{4m}{\rho^{3}}  
\end{equation}
 The above expression shows that at the conformal boundary $\rho_c \to \infty$ the first term in \eqref{Area_divergence} diverges. This divergence is precisely corrected by the singular contribution of the Gauss-Bonnet term that enters in the definition of the renormalised area.
Removing the cutoff and evaluating the renormalised area formula gives
\begin{equation}\label{Aren_Cmetric}
    \mathcal{A}_{\rm{ren}}[\Sigma_{s}]=\frac{\beta m \ell^{2}}{\alpha r^2_{+}}\left(1+s Ar_+\right)^2-2\pi\ell^2\chi(\Sigma_{s})\,.
\end{equation}
Since each worldsheet is a disk, $ \chi(\Sigma_{s})=1$. Therefore, the Euler part of each renormalised area is
\begin{equation}
    [\mathcal{A}_{\rm{ren}}(\Sigma)]_{\rm{Euler}}= -2\pi \ell^{2}\,.
\end{equation}
 Multiplying by the corresponding tensions and summing over both axes gives
 \begin{equation}\label{eq:euler-area}
     \sum_{s=\pm} \mu_{s}[\mathcal{A}_{\rm{ren}}(\Sigma_{s})]_{\rm{Euler}}=-2\pi \ell^{2} (\mu_{+}+ \mu_{-})=-2\pi \ell^{2} \mu\,.
 \end{equation}
Consequently, the finite action mismatch \eqref{eq:mismatch} satisfies
\begin{equation}
    I_{T}-I_{\rm{ren}}= - \sum_{s=\pm} \mu_{s}[\mathcal{A}_{\rm{ren}}(\Sigma_{s})]_{\rm{Euler}}\,.
\end{equation}
The same Euler contribution can be identified directly at the horizon. Let $\mathcal{H}$ denote its compact two-dimensional cross-section, with induced metric $\sigma_{ij}$ and scalar curvature $\mathscr{R}_{\mathcal{H}}$. Each string worldsheet meets $\mathcal{H}$ at a conical point with deficit fraction $\delta_{s}/2\pi = 1- a_{s}=4G\mu_{s}$. The two-dimensional Gauss--Bonnet theorem gives 
\begin{equation}
    \int\limits_{\mathcal{H}_{\rm{reg}}}d^{2}x\sqrt{\sigma}\mathscr{R}_{\mathcal{H}}+4\pi \sum_{s=\pm}\frac{\delta_{s}}{2\pi}=4\pi \chi(\mathcal{H})=8\pi\,,
\end{equation}
where $\mathcal{H}_{\rm{reg}}$ is the horizon with the two conical points excluded. Using $a_{+} + a_{-}=2/\mathcal{K}$, the integral over $\mathcal{H}_{\rm{reg}}$ gives $8\pi/\mathcal{K}$, and the two conical points contribute $8\pi(1-\mathcal{K}^{-1})$. Their sum is $8 \pi$ as required by $\chi(\mathcal{H})=2$.  Using the Gauss-Bonnet coupling $\ell^{2}/4$, the entropy contribution is \cite{Fursaev:2013fta}
\begin{equation}
    S_{GB}=\frac{\ell^{2}}{8G}\int\limits_{\mathcal{H}}d^{2}x\sqrt{\sigma}\mathscr{R}_{\mathcal{H}}=S^{\rm{reg}}_{GB}+S^{\rm{tips}}_{GB}\,,
\end{equation}
where the integral over $\mathcal{H}$ includes the curvature at the poles in the limit of a smooth regularisation. The regular and tip contributions are
\begin{equation}
S^{\rm{reg}}_{GB}= \frac{\pi\ell^{2}}{G \mathcal{K}}\,,\qquad S^{\rm{tips}}_{GB} = 2\pi \ell^{2}\mu \,.
\end{equation}
The tip contribution equals the residual mismatch between the regular action evaluations after Euler normalisation. In terms of the renormalised worldsheet areas, this gives
\begin{equation} \label{eq:final-mismatch}
\boxed{
   S^{\rm{tips}}_{GB}=I_{T}-I_{\rm{ren}}=-\sum_{s=\pm} \mu_{s}[\mathcal{A}_{\rm{ren}}(\Sigma_{s})]_{\rm{Euler}}\,}
\end{equation}


\section{Singular thermodynamics: conical deficits} \label{sec:5}
The thermodynamics of accelerating black AdS black holes with fixed conical deficits was studied in \cite{Appels:2016uha} and subsequently extended to include varying deficits in both accelerating and static backgrounds \cite{Appels:2017xoe}.  When the string tensions are allowed to vary, the first law acquires additional work terms whose conjugate quantities are known as \textit{thermodynamic lengths}. In this framework, we give a geometric interpretation of the relative thermodynamic length in terms of the difference of the renormalised worldsheet areas per unit Euclidean time. At fixed total string tension, this difference determines the work associated with redistributing tension between the axes. 
The Euclidean renormalised on-shell action satisfies the quantum statistical relation 
\begin{equation}
    I_{\rm{ren}}=\beta M - S \,, \quad S=\frac{\mathcal{A}_{\rm{H}}}{4 G}= \frac{\pi r^{2}_{+}}{G \mathcal{K}D_{+}}\,.
\end{equation}
At fixed $G$, variations within the slowly accelerating family satisfy \cite{Appels:2017xoe}
\begin{equation}
    dM = T dS + V dP - \sum_{s=\pm1}\lambda_{s}d\mu_{s}\,,\qquad P=\frac{3}{8\pi G \ell^{2}}\,,\quad T=\beta^{-1}\,,
\end{equation}
where 
\begin{equation}\label{thermo-length}
    \lambda_{s}=\frac{1}{\alpha}\left( \frac{r_{+}}{1+s Ar_{+}} -m - sA\ell^{2}\right)\,, \qquad V=
    \frac{4\pi}{3\alpha \mathcal{K}}\left(\frac{r_{+}^{3}}{D_{+}^{2}}+m A^{2}\ell^{4}\right)\,.
\end{equation}
Here, $V$ is the thermodynamic three-volume, which must not be confused with $V_{\rm{ren}}^{(E)}$, and $\lambda_s$ are the corresponding thermodynamic lengths. For the Euler normalised topological action, $I_{T}=I_{\rm{ren}}+2\pi \ell^{2}\mu$. At fixed tensions and pressure, the black hole entropy from the topological action gives
\begin{equation}
    S_{T}=(\beta\partial_{\beta}-1)I_{T}=\frac{\mathcal{A}_{\rm{H}}}{4 G}-2\pi \ell^{2}\mu\,.
\end{equation}
This shift is constant on each fixed-tension. When the total tension varies, the shift also must be varied. At fixed pressure, its contribution shifts the conjugate lengths to $\lambda^{(T)}_{s}=\lambda_{s}-2\pi \ell^{2}\beta^{-1}$.

The complete renormalised worldsheet areas contain further geometric information. The renormalised area for the C-metric \eqref{Aren_Cmetric} with \eqref{thermo-length} gives
\begin{eqnarray}
    -\frac{\mathcal{A}_{\rm{ren}}[\Sigma_{s}]}{\beta}=\lambda_{s}+\frac{m}{\alpha}(1-2 A^{2}\ell^{2})\,.
\end{eqnarray}
Multiplying by the total tension on both sides and taking its variation gives
\begin{equation}\label{tension-area}
    -\sum_{s=\pm1}\lambda_{s}d\mu_{s} =\frac{1}{\beta}\sum_{s=\pm1}\mathcal{A}_{\rm{ren}}[\Sigma_{s}]d\mu_{s}+\frac{m}{\alpha}(1-2A^{2}\ell^{2})d\mu\,.
\end{equation}
For slowly accelerating black holes, we can redistribute the tension between the two axes while keeping the total tension fixed, that is 
\begin{eqnarray}
    d\mu= d\mu_{+}+d\mu_{-}=0 \quad \Rightarrow d\mu_{-}=-d\mu_{+}\,.
\end{eqnarray}
With this, the total tension term at the right-hand side of \eqref{tension-area} vanishes. Substituting back into the first law, we get
\begin{equation} \label{eq:first-law-area}
\boxed{
    dM = \frac{T}{4G}d\mathcal{A}_{\rm{H}} + VdP + \frac{\mathcal{A}_{\rm{ren}}[\Sigma_{+}]-\mathcal{A}_{\rm{ren}}[\Sigma_{-}]}{\beta}d\mu_{+}\,.}
\end{equation}
The difference between the renormalised areas of the string worldsheet determines the relative tension work exactly. The individual thermodynamic lengths still differ from the value of the renormalised area. At fixed pressure, the volume term drops out, and the Euler mismatch $2\pi\ell^{2}\mu$ remains constant even as tension is transferred between the axes.

\section{Conclusions} \label{sec:6}

In this work, we have investigated the relation between holographic and topological renormalisation of four-dimensional AdS gravity in the presence of codimension-two conical defects. Using the slowly accelerating AdS C-metric, we have shown that the finite mismatch between the two renormalised Euclidean actions, after the standard Euler normalisation, is entirely accounted for by the tension-weighted Euler contributions of the renormalised worldsheet areas \eqref{eq:final-mismatch}. 
These contributions coincide with the Gauss--Bonnet entropy supported at the conical points of the horizon.  

In this sense, singular thermodynamics retains the finite contributions supported on the conical defects that are omitted from evaluations restricted to the regular or smooth fillings. The defect Euler contribution is constant at fixed pressure and total string tension, but varies across the broader thermodynamic family. 

We have also obtained explicit expressions for the renormalised areas of the cosmic string worldsheets. At fixed total string tension, their difference per unit Euclidean time determines the thermodynamic work associated with redistributing tension between the two axes \eqref{eq:first-law-area}. This provides a geometric interpretation of the difference of thermodynamic lengths appearing in the first law of accelerating black hole thermodynamics. 

More generally, our results illustrate how topological renormalisation naturally incorporates singular contributions that are not manifest in the standard holographic counterterm prescription. 

In a more physical context, Renormalised Volume is linked to a renormalised version of Einstein gravity action with negative cosmological constant in even dimensions, which is rendered finite by the addition of spacetime covariant terms.
Indeed, a partial renormalisation is achieved with the addition of the Euler (topological) term to the Einstein-AdS action, which,
for a metric $g_{(0)}$ that is conformally flat,  is equivalent to the counterterm method \cite{Miskovic:2009bm,Anastasiou:2020zwc}.

However, in even dimensions equal to or higher than six, the difference between Topological and Holographic Renormalisation for a generic conformal data $g_{(0)}$  starts with a Weyl-squared term at the boundary, plus higher-derivative terms constructed with conformal tensors of the boundary. Conformal Renormalisation is a way of eliminating the above drawback by the conformal completion of the Einstein-AdS action plus a topological term. This means that the regulated Einstein action in even dimensions is the second-derivative sector of a higher-derivative gravity which is conformally invariant. A nontrivial example of renormalisation of the Einstein-AdS action was given in six dimensions in \cite{Anastasiou:2020mik}. Indeed, Holographic Renormalisation appears as encoded in a precise combination of conformal invariants in the bulk. Conformal Renormalisation in eight dimensions relies on the classification of conformal invariants made in \cite{Boulanger:2004zf}. In addition, the existence of a generic Einstein sector in 8D Conformal Gravity was proven for a fine-tuned combination of these invariants  \cite{Boulanger:2025oli}. In \cite{Anastasiou:2026jrt}, the exact agreement between the Einstein sector of this particular 8D Conformal Gravity and Holographic Renormalisation was found. This argument provides further support to the idea of a deep connection between conformal invariance and finiteness of the AdS gravity action. In particular, as there is always a topological term involved in the Renormalised Volume for Einstein spaces, it is expected that one could readily capture codimension-two defects of gravitational solutions (and singular thermodynamic contributions) in this way. 

\paragraph{Note added.} While this work was being completed, a manuscript with conceptual overlap regarding the interpretation of finite topological contributions to the Euclidean action and black hole entropy appeared on arXiv \cite{Liska:2026srl}. 

\section*{Acknowledgments}
 We thank Felipe Diaz for his collaboration during the initial stage of this work. We also thank Alexey Koshelev and Anna Tokareva for their valuable comments and discussions on related topics. The work of GAH was supported by the National Natural Science Foundation of China (NSFC) grant No. 12505091. This work was supported by the Agencia Nacional de Investigación y Desarrollo (ANID) through FONDECYT Regular grants No.~1230492, 1231779, 1240043, 1240955, 1250672, and 1261016.

\bibliographystyle{JHEP}
\bibliography{ref}
\end{document}